\documentclass[apj]{emulateapj}
\usepackage{natbib}
\usepackage{amsmath}
\usepackage{amssymb}
\usepackage{graphicx}
\usepackage[flushleft]{threeparttable}
\usepackage{bm}
\shorttitle{X-ray emissivity of old stellar populations}
\shortauthors{Chong Ge et al.}

\begin{document}
\title{X-ray emissivity of old stellar populations: a Local Group census}
\author
{Chong Ge$^{1,2,3,4}$, Zhiyuan Li$^{1,2,3}$, Xiaojie Xu$^{1,2,3}$, Qiusheng Gu$^{1,2,3}$, Q. Daniel Wang$^{1,4}$, 
Shawn Roberts$^{4}$,\\
Ralph P. Kraft$^{5}$,
Christine Jones$^{5}$,
and William R. Forman$^{5}$}
\affil{$^{1}$ School of Astronomy and Space Science, Nanjing University, Nanjing 210093, China}
\affil{$^{2}$ Key Laboratory of Modern Astronomy and Astrophysics (Nanjing University), Ministry of Education, Nanjing 210093, China}
\affil{$^{3}$ Collaborative Innovation Center of Modern Astronomy and Space Exploration, Nanjing 210093, China}
\affil{$^{4}$ Department of Astronomy, University of Massachusetts, Amherst, MA 01003, USA}
\affil{$^{5}$ Harvard-Smithsonian Center for Astrophysics, 60 Garden Street, Cambridge, MA 02138, USA}

\email{lizy@nju.edu.cn} 

\begin{abstract}
We study the unresolved X-ray emission in three Local Group dwarf elliptical galaxies (NGC\,147, NGC\,185 and NGC\,205) using {\sl XMM-Newton} observations, which most likely originates from a collection of weak X-ray sources, mainly cataclysmic variables and coronally active binaries. Precise knowledge of this stellar X-ray emission is crucial not only for understanding the relevant stellar astrophysics but also for disentangling and quantifying the thermal emission from diffuse hot gas in nearby galaxies.
We find that the integrated X-ray emissivities of the individual dwarf ellipticals agree well with that of the Solar vicinity, supporting an often assumed but untested view that the X-ray emissivity of old stellar populations is quasi-universal in normal galactic environments, in which dynamical effects on the formation and destruction of binary systems are not important.
The average X-ray emissivity of the dwarf ellipticals, including M32 studied in the literature, is measured to be $L_{0.5-2\ \rm {keV}}/M_{\ast} = (6.0 \pm 0.5 \pm 1.8) \times 10^{27} \ \rm{erg \ s^{-1} \ M_\odot^{-1}}$. 
We also compare this value to the integrated X-ray emissivities of Galactic globular clusters and old open clusters and discuss the role of dynamical effects in these dense stellar systems.
\end{abstract}


\section{Introduction}
X-ray studies of nearby normal galaxies (without an active nucleus) have established two primary components of the X-ray emission: discrete sources and diffuse hot gas.
The discrete sources include low- and high-mass X-ray binaries (LMXBs and HMXBs) with $L_X \gtrsim 10^{35}-10^{36} \ \rm{erg \ s^{-1}}$. 
In early-type galaxies, except gas-rich, massive ellipticals, the overall X-ray luminosity is dominated by LMXBs, the integrated luminosity of which scales primarily with stellar mass (Gilfanov 2004) but also depends on stellar age (Zhang, Gilfanov \& Bogd{\'a}n 2012).
In late-type galaxies, apart from LMXBs, HMXBs contribute to the bulk X-ray emission, whose total luminosity is roughly proportional to the star formation rate (Grimm, Gilfanov \& Sunyaev 2003).
The diffuse X-ray emission from hot gas traces the feedback from both young and old stellar populations (core-collapsed and Type Ia SNe, and winds from evolved stars) and activity of the central super-massive black hole,
which are believed to play a critical role in regulating the energy, mass, and metal transport, as well as the dynamics of the interstellar medium (e.g., Strickland et al. 2004; David et al. 2006; Li, Wang \& Hameed 2007; Fraternali \& Binney 2008; Tang et al. 2009a,b).
Studying the thermodynamic properties of hot gas in and around galaxies is crucial to our fundamental understanding of galaxy formation and evolution.

However, studies of hot gas in most nearby galaxies have been challenging even with the bright sources detected and removed, because the seemingly diffuse X-ray emission must contain, and in some cases be dominated by, unresolved emission from faint stellar sources, whose contributions have not been carefully quantified until recently. These sources, mainly cataclysmic variables (CVs) and coronally active binaries (ABs) with $L_{X} \sim 10^{28}-10^{34} \ \rm{erg \ s^{-1}}$, are shown to be responsible for the unresolved 
X-ray emission distributed over the inner region of our Galaxy, the so-called Galactic Ridge X-ray Emission (Revnivtsev et al. 2006).
Studies of the hot gas thus rely on a proper isolation of this stellar component, especially in the soft X-ray ($0.5-2$ keV) band, where the truly diffuse emission concentrates.
However, since the X-ray emission of CVs and ABs essentially arises from optically thin thermal plasmas with temperatures similar to that of the diffuse hot gas, it is not straightforward to separate the stellar and gaseous contributions spectroscopically, albeit the emission from CVs can be relatively harder (e.g. Heinke et al. 2008). 
On the other hand, one may expect that the stellar X-ray emission is approximately linearly proportional to the stellar mass of the host galaxy.

Previous studies on the Solar vicinity (Sazonov et al. 2006), the compact dwarf elliptical (dE) galaxy M32 (Revnivtsev et al. 2007), the intermediate-mass, gas-poor elliptical galaxy NGC\,3379 (Revnivtsev et al. 2008), and the outer bulge of the early-type spiral galaxy M31 (Li \& Wang 2007; Bogd{\'a}n \& Gilfanov 2008) showed that the cumulative X-ray luminosities per unit stellar mass (i.e., stellar X-ray emissivity) from these galaxies/galactic regions
agree within a factor of 2 with each other, suggesting a quasi-universal stellar X-ray emissivity (Revnivtsev et al. 2008).
However, the number of CVs and ABs found in the Solar neighborhood is small and might be peculiar; NGC\,3379 may contain a certain amount of hot gas given its moderate gravitational mass (Trinchieri et al. 2008; Bogd{\'a}n et al. 2012);
the outer bulge of M31 is also likely to harbor some hot gas, although its contribution to the unresolved X-ray emission is small (Li \& Wang 2007).
Perhaps M32 is the only \emph{bona fide} gas-free host allowing for an accurate measurement of the stellar X-ray emissivity.
Indeed, the unresolved X-ray emission from M32 has been widely adopted in recent studies (e.g., Boroson, Kim \& Fabbiano 2011; Li \& Wang 2013) to account for the contribution of CVs and ABs in nearby galaxies, after scaling with the stellar mass.
Nevertheless, whether a quasi-universal stellar X-ray emissivity exists remains an open question.

In this work we perform an X-ray study of three nearby dEs, NGC\,147, NGC\,185 and NGC\,205, in order to advance our knowledge of the stellar X-ray emissivity. These dEs are the best suitable targets in the Local Group to measure the cumulative X-ray emission of old stellar populations, since 
i) they are not expected to confine a significant amount of hot gas, given their shallow gravitational potential;
ii) they are not likely to have recent star formation and are therefore free of HMXBs, SNRs, and pre-main sequence stellar objects that might contribute to the X-ray emission;
and iii) they are however among the most massive dwarf galaxies in the Local Group and are located at appropriate distances so that good signal-to-noise ratios (S/N) are warranted and that the contribution of any bright LMXBs can be properly subtracted.

For comparison, we also study four Galactic globular clusters (GCs), $\omega$ Cen (Haggard et al. 2009), 47 Tuc (Heinke et al. 2005), NGC\,6266 and NGC\,6397 (Bogdanov et al. 2010), 
which are among the X-ray brightest GCs and all have high quality X-ray data. 
Compared to the typical stellar density of $\rm\bm{0.1-1\ M_{\odot}\ pc^{-3}}$ in the galactic fields (e.g., in the Solar vicinity and in the dEs), 
the stellar density in the core of GCs can reach $\gtrsim \rm\bm{10^4\ M_{\odot}\ pc^{-3}}$ (Binney \& Tremaine 1987), leading to an enhanced stellar encounter rate 
that may affect the number of X-ray sources (Pooley et al. 2003; Pooley 2010).
The majority of X-ray sources in GCs are expected to be CVs and ABs (Heinke 2010), 
allowing for a direct comparison between their integrated X-ray emission and the unresolved X-ray emission from the dEs. 
 
The remainder of the paper is organized as follows:
Section 2 describes the {\sl XMM-Newton} and {\sl Chandra} observations and data reduction;
Section 3 presents our data analysis and results;
Section 4 compares the dEs, GCs and other old stellar systems in the scope of a quasi-universality of stellar X-ray emissivity and dynamical effects;
Section 5 summarizes our findings.

\section{Observation and data reduction}
\subsection{XMM-Newton}
We utilized 4 {\sl XMM-Newton} observations on the three dEs. Among these, the deepest observation is Obs. ID 0652210101 (PI: Z.Li) for NGC\,185, which has an exposure of $\sim$115 ks ($\sim$85 ks after filtering the time intervals of high particle background; Table 1), while other observations are from the {\sl XMM-Newton} archive.  
We used the Science Analysis System (SAS, version 13.0.0), together with the corresponding calibration files for data reduction, following the procedures in Li et al. (2006).
Point sources were detected in the $0.5-7$ keV band and removed in the subsequent analysis. The spectrum of the unresolved emission after removing the detected sources was extracted from the half-light ellipse (with Ks-band major-to-minor axis ratio and position angle from the 2MASS Large Galaxy Atlas\footnote{http://www.ipac.caltech.edu/2mass/}; Jarrett et al. 2003).
The background spectrum was extracted from elliptical annuli surrounding the source region (Table 2). 

\subsection{Chandra}
We analyzed 9 archival {\sl Chandra} observations on the four GCs.
We reprocessed the data using the {\sl Chandra} Interactive Analysis of Observations (CIAO, version 4.5) software, following the procedures in Li et al. (2010).
Point sources were detected in the $0.3-7$ ($0.5-8$) keV for the ACIS-S (ACIS-I) data. The source detection limits are listed in Column (8) of Table 2.
We found no luminous LMXBs in these GCs; all detected sources have luminosity $L_{0.5-7\ \rm {keV}} < 10^{34} \ \rm{erg \ s^{-1}}$, which are in agreement with previous work (Heinke et al. 2005, Haggard et al. 2009, Bogdanov et al. 2010). 
We further examined the existence of bright sources that may dominate the total X-ray emission. The brightest source in 47 Tuc contributes $\sim$32\% of the cluster's total X-ray flux, which was suggested to be a qLMXB (Heinke et al. 2005). However, the luminosity of this source lies within the range of typical CVs and ABs, and would be unresolved if located in the dEs. Therefore we retained this source for subsequent analysis. For each GC, the accumulated spectrum was extracted from within the half-light circle including detected sources and the unresolved emission,
while the background spectrum was extracted from an annulus surrounding the source region and excluding detected point sources therein.
Table 1 gives a log of the {\sl XMM-Newton} and {\sl Chandra} observations.

\begin{table}
 \centering
  \caption{Log of {\sl XMM-Newton}/{\sl Chandra} observations}
  \scriptsize
  \begin{tabular}{@{}lccc@{}}
\hline 
Name & Obs-ID & Exp(ks) & CleanExp(ks)\\ 
\hline 
NGC\,147 & 0204790201 (X)$^a$ & 13.5 & 7.2/11.1/10.6$^b$\\
NGC\,185 & 0204790301, 0652210101 (X) & 125.5 & 66.4/86.8/85.8\\
NGC\,205 & 0204790401 (X) & 18.4 & 6.7/7.0/7.0\\
\hline
$\omega$ Cen & 653, 1519 (C)  & 69.2 & 63.7\\
47 Tuc & 2735, 2736, 2737, 2738 (C) & 254.3 & 246.3\\
NGC\,6266 & 2677(C)  & 61.7 & 59.5\\
NGC\,6397 & 7460, 7461 (C) & 237.4 & 230.3\\ 
M32 & 2017, 2494, 5690 (C) & 174.9 & 158.9 \\ 
\hline 
\end{tabular}
$^a$X and C represent {\sl XMM-Newton} and {\sl Chandra}, respectively.\\
$^b$Clean exposures of the pn/MOS1/MOS2 cameras.
\end{table}

\section{Data analysis and results}
\subsection{Dwarf elliptical galaxies}

\begin{figure}
\begin{center}
\includegraphics[width=8.0cm,angle=0]{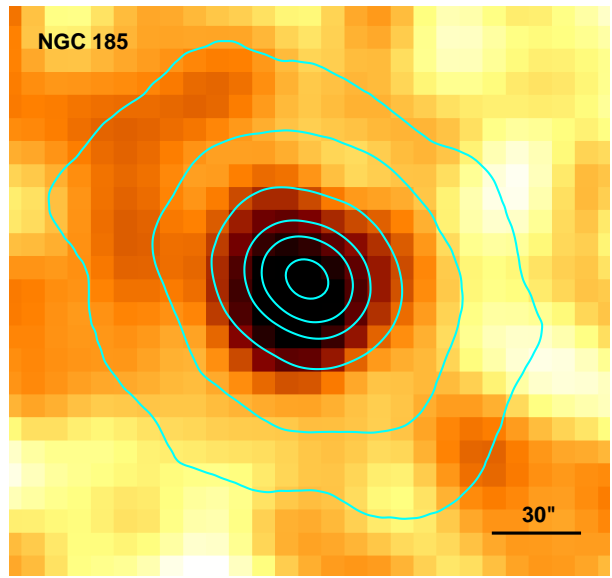}
\vskip 3mm
\caption{The combined {\sl XMM-Newton} pn, MOS1 and MOS2 image (logarithmically scaled and smoothed with a 20$^{\prime\prime}$ Gaussian kernel) of the $0.5-2$ keV unresolved emission in NGC\,185, overlaid by 2MASS Ks-band intensity contours. The scale bar in the image is 30 arcsec, corresponding to a linear scale of $\sim$90 pc.}
\label{f:N185image}
\end{center}
\end{figure}

\begin{figure}
\centering
\includegraphics[width=8.0cm, angle=0]{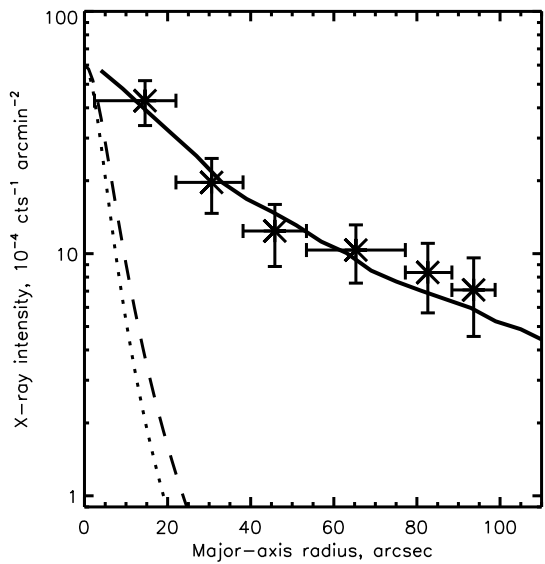}
\caption{$0.5-2$ keV radial intensity profile (from the combined pn, MOS1 and MOS2 image) of NGC\,185, which is adaptively binned to achieve a S/N better than 5. The intensity is computed for elliptical annuli with major-to-minor axis ratio of 0.87 and position angle of 52$^\circ$ (east from north). The solid line is the 2MASS Ks-band intensity profile tracing the starlight. The dotted and dashed lines represent the PSF of MOS1 and pn, respectively. }
\label{f:sbd}
\end{figure}

We first examined NGC\,185, which has the highest S/N. Fig.~\ref{f:N185image} shows the {\sl XMM-Newton} image in the $0.5-2$ keV band, overlaid by 2MASS Ks-band intensity contours tracing the starlight. In Fig.~\ref{f:sbd} we compared the radial intensity profile of the unresolved X-ray emission with the Ks-band profile. Also plotted are the MOS1 point-spread function (PSF) as dotted line and the pn PSF as dashed line (Ghizzardi 2002). It is evident that the X-ray intensity profile differs from a point source, which indicates that the X-ray emission is truly extended, rather than dominated by a central, compact source. The similarities in the morphology and radial intensity distribution between the X-rays and near-infrared starlight strongly suggest a stellar origin for the unresolved X-ray emission.

\begin{figure}
\centering
\includegraphics[width=0.5\textwidth, angle=0]{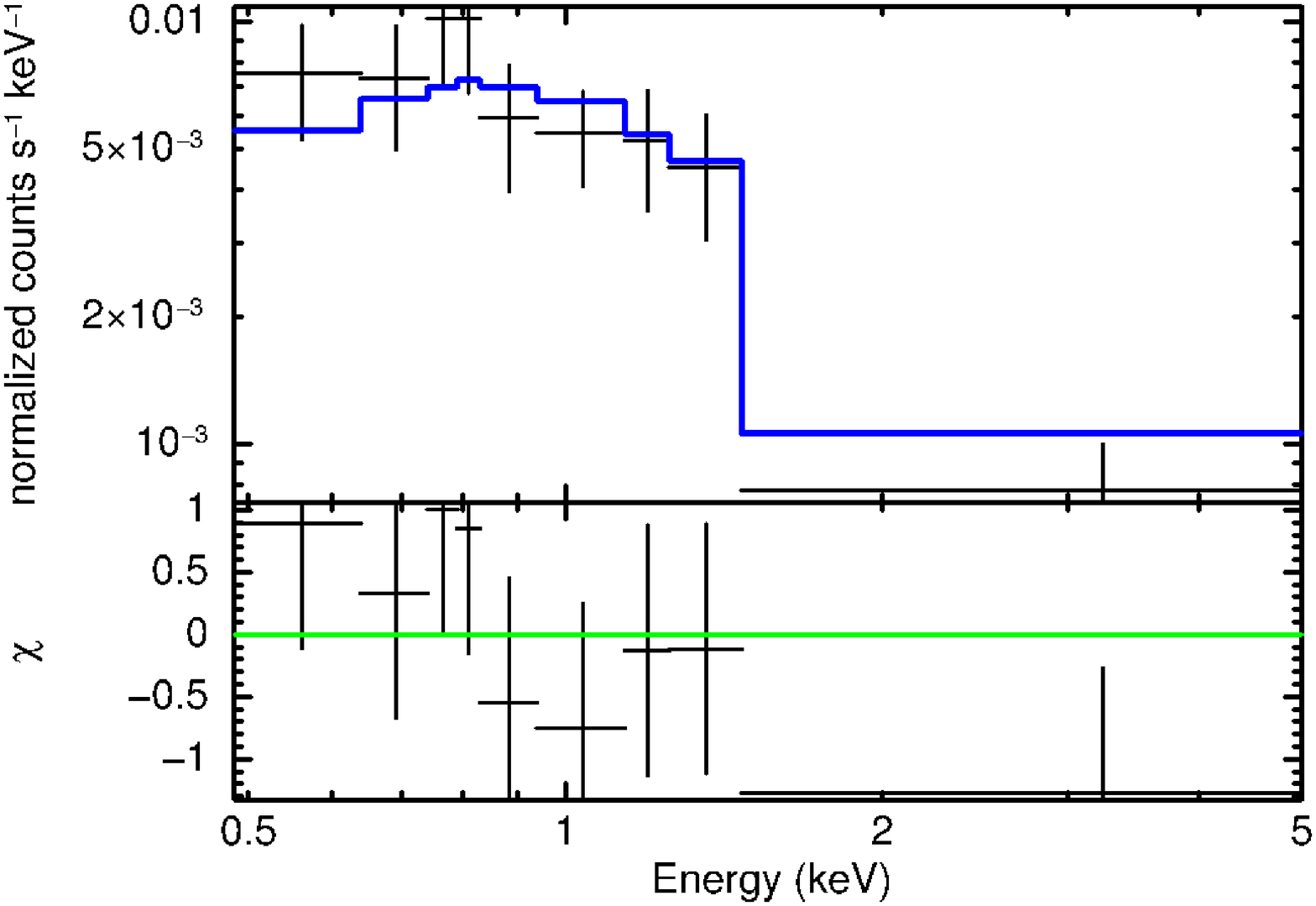}
\caption{{\sl XMM-Newton} spectrum of the unresolved emission of NGC\,185, which is adaptively binned to achieve a S/N better than 3. The blue histogram is the best-fit absorbed log-normal plasma temperature distribution model.}
\label{f:N185spec}
\end{figure}

To fit the X-ray spectrum of NGC\,185, we used an absorbed log-normal plasma temperature distribution model, which is discussed in the Appendix. The idea of using this model is invoked by the underlying multi-temperature nature of the stellar X-ray emission. 
In the fit, we fixed the metallicity ([Fe/H]=-1.3 from McConnachie 2012) and adopted the solar abundance standard of Anders \& Grevesse (1989), thus the model has three free parameters: peak temperature ($T_{peak}$), dispersion ($\sigma_T$) and normalization. 
Fig.~\ref{f:N185spec} shows the NGC\,185 spectrum and its best-fit model, with $\rm{\chi^{2}_{red}/dof}$ = 1.03/5, $kT_{peak} = 0.9 \pm 0.6$ keV and $\sigma_T$ = 3.0 $\pm$ 1.0. Here and below errors in the spectral fitting results are given at the 90\% confidence level.

\begin{figure}
\centering
\includegraphics[width=0.5\textwidth, angle=0]{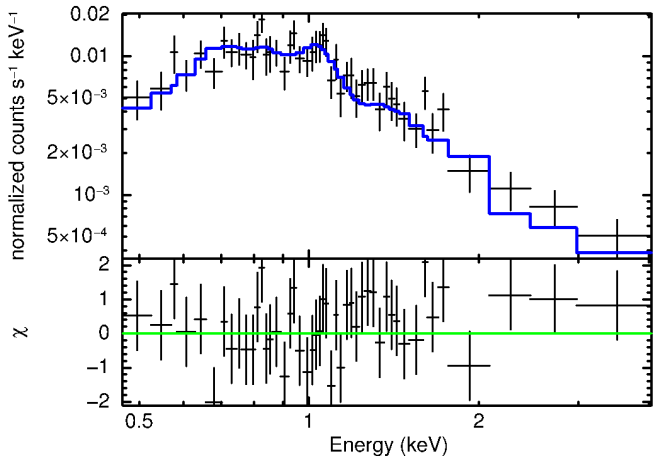}
\caption{{\sl Chandra} spectrum of the unresolved emission of M32, 
which is adaptively binned to achieve a S/N better than 3. The blue histogram is the best-fit absorbed log-normal plasma temperature distribution model.}
\label{f:M32spec}
\end{figure}

For NGC\,147 and NGC\,205, their spectral properties cannot be well constrained due to the limited S/N (Table 1). We thus made use of the {\sl Chandra} spectrum of M32 (Li \& Wang 2007), which is dominated by CVs and ABs and has a much better S/N. We adopted the log-normal model 
to fit this spectrum with a fixed metallicity ([Fe/H]=-0.25 from McConnachie 2012). The result is shown in Fig.~\ref{f:M32spec} with best model parameters as $\rm{\chi^{2}_{red}/dof}$ = 0.91/45, $kT_{peak} = 1.3 \pm 0.2$ keV and $\sigma_T$ = 3.0 $\pm$ 0.3. 
We assumed that the spectra of dEs are similar with each other, because they share a common stellar origin.
To test this assumption, we fit the spectrum of NGC\,185 with the best-fit model of M32, leaving only the normalization as free parameter. 
The resultant flux is different from the above independent fit by only 6$\%$. 
We then used the best-fit model of M32 to obtain the unabsorbed fluxes of NGC\,147 and NGC\,205 (Table 2).

Next, we used the 2MASS Ks-band images to estimate the near-infrared luminosity and stellar mass of the dEs. We converted the integrated, 
background-subtracted Ks-band flux from within the half-light ellipse to the luminosity, using distances from McConnachie (2012) and correcting for the Galactic foreground extinction (Dickey \& Lockman 1990). 
The stellar mass was estimated by adopting the color-dependent ($B-V$; from NED\footnote{http://ned.ipac.caltech.edu/}) mass-to-light ratio ($\sim$0.8) from Bell et al. (2003). 
We then calculated the 0.5-2 keV and 2-8 keV emissivities for each dE, dividing the model-predicted X-ray luminosity by the stellar mass. The results are given in Table 2, in which a 30\% systematic uncertainty
in the estimated stellar mass (Bell et al. 2003) has been included.

For the Galactic GCs, one can in principle detect any bright LMXBs if they exist (Section 2.2), but in the case of the more distant dEs, due to the inferior source detection limit, unresolved LMXBs may contribute to the total X-ray emission.
We integrated the luminosity function of LMXBs derived from nearby galaxies (Gilfanov 2004) from the source detection limit of individual dEs (Column 8 in Table 2) down to $10^{34} \ \rm{erg \ s^{-1}}$ to calculate the total $0.5-2$ keV emissivity from unresolved LMXBs, assuming a typical power-law spectrum with photon-index $\Gamma$ = 1.56 (Irwin, Athey \& Bregman 2003).
For NGC\,147, NGC\,185 and NGC\,205, the contributions from unresolved LMXBs were found to be 2.2\%, 0.7\% and 3.8\%, respectively, which are negligible.

\newpage

\subsection{Globular clusters}

For the four GCs, we also adopted an absorbed log-normal temperature model to fit their spectra, with the metallicities fixed at the values listed in the Globular Cluster Catalog of Harris (2010). 
The best-fit parameters for the GCs as well as for NGC\,185 and M32 are summarized in Table 3.
The fit was acceptable for $\omega$ Cen, NGC\,6266 and NGC\,6397, with a somewhat broader temperature range than in the dEs. 
For 47 Tuc, the log-normal temperature model was able to characterize the overall shape of the spectrum but resulted in sinusoidal residuals. Thus 
we have also tried a two-temperature model, which led to an improved fit, with best-fit temparures of $0.23\pm0.01$ and $3.8\pm0.2$ keV. However, the model-predicted fluxes differ by less than 5\% between the log-normal temperature and two-temperature models\footnote{This is also the case for the spectral fit of the other GCs and the dEs.}.
We then adopted the photometric measurements from Harris (2010) to estimate the total stellar masses of the GCs, using a fixed mass-to-light ratio of $M/L_V = 3$ (Gnedin et al. 2002; Harris et al. 2006). Simple
Stellar Population models (e.g., Bruzual \& Charlot 2003; Anders \& Fritze-v. Alvensleben 2003) predict  $M/L_V = 2-4$ for typical GC metallicities (Kruijssen 2008). Therefore, we accounted for a 33\% systematic error in the estimated GC masses.
Finally we derived the cumulative X-ray emissivities of the GCs (Table 2).

Recently, Wu et al.~(2014) claimed the presence of truly diffuse X-ray emission in the core of 47 Tuc. We extracted a spectrum of the unresolved emission (i.e., masking all detected sources) 
within the half-light radius of 47 Tuc, and found that the unresolved emission contributes only 24\% of the total flux. Thus any truly diffuse emission
does not significantly affect our results.

We have also estimated possible contamination of background AGNs with the cumulative flux distribution of Brandt et al. (2001), assuming a power-law spectrum with photon-index of 1.4. We found that background AGNs may contribute up to 44\%, 11\%, 1\% and 10\% of the cumulative flux in $\omega$ Cen, 47 Tuc, NGC\,6266 and NGC\,6397, respectively, but our conclusions below remain unchanged. 

\setlength{\tabcolsep}{1pt}
\begin{table*}
\begin{threeparttable}
\caption{Physical properties of the dwarf elliptical galaxies and globular clusters}
\tabcolsep=0.08cm
\begin{tabular}{@{}lccccccccc}
\\
\hline 
Name & Distance & R$_e$ & N$_{\rm H}$ & [Fe/H] &  L &  M  & Detection limit & 0.5-2 keV emissivity & 2-8 keV emissivity\\
 & (kpc) & (arcmin) & ($10^{20} {\rm~cm^{-2}}$) &  & (L$_\odot$) &   (M$_\odot$) & ($\rm{erg \ s^{-1}}$) & ($10^{27} \ \rm{erg \ s^{-1} \ M_\odot^{-1}}$) & ($10^{27} \ \rm{erg \ s^{-1} \ M_\odot^{-1}}$)\\ 
 (1) & (2) & (3) & (4) & (5) & (6) & (7) & (8) & (9) & (10) \\
 \hline 
 NGC\,147 & 676 & 2.27/0.49/52 & 12.1 & -1.1 & $1.3\times10^8$ & $1.1\times10^8$ & $7.8 \times 10^{35}$ & $11.2 \pm 2.9 \pm 3.5$ & $13.3 \pm 3.6 \pm 4.0$\\  
 NGC\,185 & 617 & 1.63/0.87/52 & 12.4 & -1.3 & $2.0\times10^8$ & $1.6\times10^8$ & $1.4 \times 10^{35}$ & $6.1 \pm 0.7 \pm 1.8$ & $6.3 \pm 0.7 \pm 1.9$\\ 
 NGC\,205 & 824 & 2.97/0.62/177 & 6.7 & -0.8 & $8.5\times10^8$ & $6.8\times10^8$ & $8.9 \times 10^{35}$ & $7.3 \pm 1.6 \pm 2.2$ & $7.8 \pm 1.8 \pm 2.3$ \\
 \hline 
 $\omega$ Cen & 5.2 & 5.00 & 6.7 & -1.53 & $1.1\times10^6$ &$3.3\times10^6$ & $1.9 \times 10^{30}$ & $0.7 \pm 0.1 \pm 0.2$ & $1.1 \pm 0.1 \pm 0.4$\\
 47 Tuc & 4.5 & 3.17 & 2.2 & -0.72 & $5.0\times10^5$ &$1.5\times10^6$ & $9.2 \times 10^{29}$ & $3.9 \pm 0.1 \pm 1.3$ & $4.1 \pm 0.1 \pm 1.4$\\
 NGC\,6266 & 6.8 & 0.92 & 26.1 & -1.18 & $4.0\times10^5$ & $1.2\times10^6$ & $3.0 \times 10^{30}$ & $5.1 \pm 0.3 \pm 1.7$ & $6.0 \pm 0.4 \pm 2.0$\\
 NGC\,6397 & 2.3 & 2.90 & 9.9 & -2.02 & $3.9\times10^4$ & $1.2\times10^5$ & $3.8 \times 10^{29}$ & $7.8 \pm 0.2 \pm 2.6$ & $20.3 \pm 0.5 \pm 6.7$\\
\hline 
\end{tabular}
\begin{tablenotes}
      \small
\item \textit{Note.} (1) Object names. (2) Distances of dEs from McConnachie (2012) and of GCs from Harris (2010).
(3) The Ks-band major axis/axis ratio/position angle of the half-light ellipse from the 2MASS Large Galaxy Atlas catalog (Jarrett et al. 2003) for dEs; half-light radius from Harris (2010) for GCs. 
(4) Galactic foreground absorption column density from Dickey \& Lockman (1990) for dEs; conversion following Predehl \& Schmitt (1995) with E(B-V) taken from Harris (2010) for GCs.
(5) Metallicity from McConnachie (2012) for dEs and from Harris (2010) for GCs, respectively.
(6) Ks-band total luminosities from 2MASS for dEs and V-band total luminosities from Harris (2010) for GCs.
(7) For dEs, the stellar mass was measured from Ks-band luminosity by applying the color-dependent ($B-V$) mass-to-light ratio $\sim$0.8, with a systematic uncertainty of $\sim$30\% (Bell et al. 2003); for GCs, the stellar mass was estimated from V-band luminosity with $M/L_V = 3$ (Gnedin et al. 2002; Harris et al. 2006) with an uncertainty of $\sim$33\% predicted by Simple
Stellar Population models (e.g., Bruzual \& Charlot 2003; Anders \& Fritze-v. Alvensleben 2003) for typical GC metallicities. We divided these masses by a factor of 2 when calculating the emissivities.
(8) Source detection limit in $0.5-7$ keV for dEs, and in $0.3-7/0.5-8$ keV for ACIS-S (47 Tuc, NGC\,6266, NGC\,6397)/ACIS-I ($\omega$ Cen) data. A power-law model with $\Gamma$ = 1.56 (Irwin, Athey \& Bregman 2003) was used to compute the values.
(9) $0.5-2$ keV and (10) 2-8 keV luminosities per stellar mass. The first and second error terms give the uncertainties in the X-ray flux and the stellar mass, respectively.
    \end{tablenotes}
  \end{threeparttable}
\end{table*}


\begin{table}
 \centering
  \caption{The Log-Normal Temperature spectral model fit for the dwarf ellipticals and globular clusters}
  \scriptsize
  \begin{tabular}{@{}lcccc@{}}
\hline 
Name & $\rm{\chi^{2}_{red}/dof}$ & $kT_{peak}$ & $\sigma_T$ & 0.5-2 keV Flux\\ 
& &  (keV) & & ($10^{-14}\rm{~erg~s^{-1}~cm^{-2}}$)\\
 \hline
M32  & 0.91/45 & $1.3 \pm 0.2$ & $3.0\pm0.3$ & $3.9 \pm 0.2$\\
NGC\,185 & 1.03/5 & $0.9 \pm 0.6$ & $3.0\pm1.0$ & $1.0 \pm 0.1$\\
\hline
$\omega$ Cen  & 1.04/49 & $3.9 \pm 0.3$ & $7.5\pm0.7$ & $34.2 \pm 2.8$\\
47 Tuc &  3.38/289 & $1.4 \pm 0.1$ & $4.2 \pm 0.1$ & $119.5 \pm 1.0$\\
NGC\,6266 & 1.17/179 & $0.6 \pm 0.1$ & $8.4 \pm 1.0$ & $55.4 \pm 3.5$\\
NGC\,6397  & 1.18/299 & $0.2 \pm 0.1$ & $18.2 \pm 0.1$ & $74.0 \pm 1.9$\\ 
\hline 
\end{tabular}
\end{table}

\begin{figure*}
\centering
\includegraphics[width=14.0cm, angle=0]{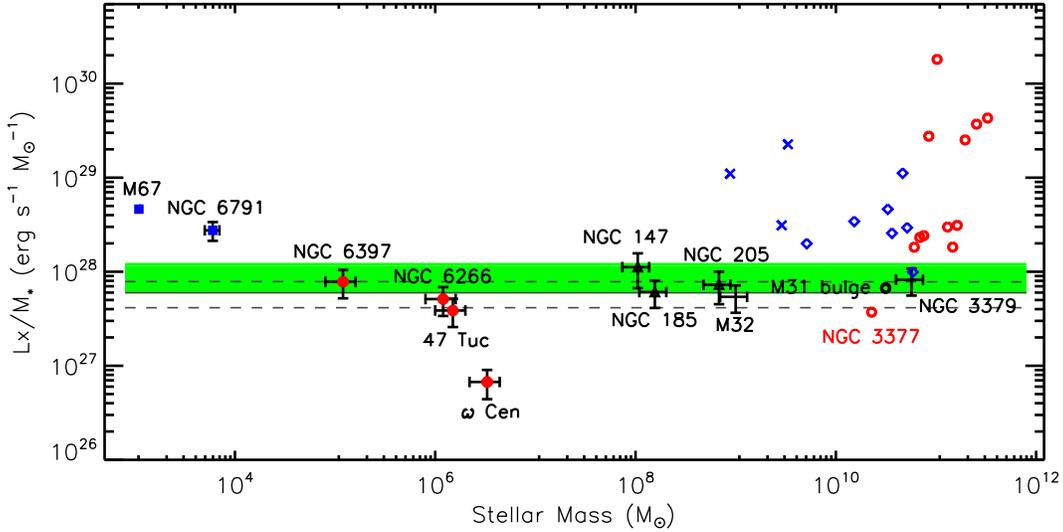}
\caption{
$0.5-2$ keV emissivities of galaxies and star clusters: 
M32 (Revnivtsev et al. 2007) and NGC\,3379 (Revnivtsev et al. 2008) are in black `+';
12 early-type galaxies (red open circles), the outer bulge of M31 (black open circle), 7 spiral galaxies (diamonds) and 3 irregular galaxies (crosses) are from Bogd{\'a}n \& Gilfanov (2011); 
also denoted are the three dEs (NGC\,147, NGC\,185 and NGC\,205) in triangles,  
the four GCs ($\omega$ Cen, 47 Tuc, NGC\,6266 and NGC\,6397) in filled circles and two old OCs (M67 and NGC\,6791; van den Berg et al. 2013) in filled squares.
The green strip represents the cumulative emissivity of ABs and CVs in the Solar vicinity (Sazonov et al. 2006), whereas the pair of dashed lines mark the range
of the dEs (Equation 2).
The error bars for the dEs and GCs include statistical and systematic errors. Errors on the individual galaxies from Bogd{\'a}n \& Gilfanov (2011) are neglected, but their stellar masses
are in principle also subject to $\sim$30\% systematic uncertainties, and their X-ray fluxes were derived with a different spectral model.
}
\label{f:emissivity}
\end{figure*}

\section{Discussion}

\subsection{A quasi-universal X-ray emissivity of old stellar populations}
The above results for dEs and GCs are summarized in Fig.~\ref{f:emissivity}  
which shows the soft X-ray ($0.5-2$ keV) emissivity versus the underlying stellar mass. For comparison, we also include literature studies on the Solar vicinity (Sazonov et al. 2006), various nearby galaxies (Revnivtsev et al. 2007, 2008; Bogd{\'a}n \& Gilfanov 2011) and two Galactic old open clusters (OCs; van den Berg et al. 2013). 

For the galaxies, the contributions from bright X-ray sources, such as LMXBs and HMXBs, have been excluded.
We emphasize that the importance of our sample of dEs lies in their low stellar masses ($ < 10^9M_{\odot}$) and the probable absence of hot gas and young stellar populations (Davidge 2005). 
Compared to the dEs, the massive early-type galaxies (red open circles in Fig.~5) show substantially higher soft X-ray emissivity, owing to the presence of a large amount of hot gas that often dominates the diffuse emission (Mathews \& Brighenti 2003). 
The only exception is NGC\,3377, which may be due to over-subtraction of the unresolved LMXB contribution, according to Bogd{\'a}n \& Gilfanov (2011). Both NGC\,3379 and the outer bulge of M31 show similar emissivities with the dEs, indicating that they harbor little, if any, truly diffuse hot gas (Bogd{\'a}n \& Gilfanov 2011; Bogd{\'a}n et al. 2012).
 The irregular galaxies and most spiral galaxies (blue crosses and diamonds in Fig.~5) in the Bogd{\'a}n \& Gilfanov (2011) sample also exhibit higher emissivities, because part of their unresolved X-ray emission should originate from young stellar populations, and they also contain a significant amount of hot gas (Li \& Wang 2013). 

On the other hand, the X-ray emissivities of the dEs are consistent with the cumulative emissivity ($9 \pm 3 \times 10^{27} \ \rm{erg \ s^{-1} \ M_\odot^{-1}}$; green strip in Fig.~5) of ABs and CVs 
in a normal galactic environment, i.e., with stellar densities typical of the field.
We calibrate this stellar X-ray emissivity by averaging over the four Local Group dEs (NGC\,147, NGC\,185, NGC\,205
and M32\footnote{The emissivity of M32 is from Revnivtsev et al. (2007): $L_{0.5-2\ \rm {keV}}/M_{\ast} = (5.4 \pm 0.7 \pm 1.6) \times 10^{27} \ \rm{erg \ s^{-1} \ M_\odot^{-1}}$, and $L_{2-8\ \rm {keV}}/M_{\ast} = (3.7 \pm 1.1 \pm 1.4) \times 10^{27} \ \rm{erg \ s^{-1} \ M_\odot^{-1}}$, which is converted from their $2-7$ keV emissivity through the best-fit model of M32 (Section 3.1).}), which results in,
\begin{equation}
L_{0.5-2\ \rm {keV}}/L_{K} = (4.7 \pm 0.4) \times 10^{27} \ \rm{erg \ s^{-1} \ L_\odot^{-1}},
\end{equation} 
\begin{equation}
L_{0.5-2\ \rm {keV}}/M_{\ast} = (6.0 \pm 0.5 \pm 1.8) \times 10^{27} \ \rm{erg \ s^{-1} M_\odot^{-1}},
\end{equation} 
\begin{equation}
L_{2-8\ \rm {keV}}/L_{K} = (4.6 \pm 0.4) \times 10^{27} \ \rm{erg \ s^{-1} \ L_\odot^{-1}},
\end{equation}
\begin{equation}
L_{2-8\ \rm {keV}}/M_{\ast} = (5.9 \pm 0.6 \pm 1.8) \times 10^{27} \ \rm{erg \ s^{-1} \ M_\odot^{-1}},
\end{equation} 
in which the first error term gives the statistical uncertainty in the X-ray flux measurement, and the second term accounts for the $\sim$30\% systematic uncertainty in the estimated stellar mass.

\subsection{The influence of a dense stellar environment}
In Fig.~5, the two old OCs, M67 and NGC\,6791 (van den Berg et al. 2013)\footnote{We converted their $0.3-7$ keV emissivities into the $0.5-2$ keV band through a 2 keV thermal plasma model.}, show substantially higher emissivities than the dEs and the Solar vicinity,  as well as the emissivities of the four GCs. The latter trend was also noted by Verbunt (2001; Figure 7 therein) based on ROSAT observations of a sizable sample of GCs.
As in GCs, the total X-ray emission of OCs is thought to be dominated by CVs and ABs (van den Berg et al. 2013), thus their higher X-ray emissivity
may be attributed to a higher binary fraction that resulted from such dynamical effects as mass segregation and evaporation (Portegies Zwart et al. 2001). Mass segregation causes heavy binaries to congregate at the cluster core, while the lighter single stars are expelled towards the periphery. Evaporation causes high velocity, usually single stars to escape, resulting in a reduced cluster mass.
We note, however, accurate measurements of binary fraction exist for only few OCs to date (e.g., Sollima et al. 2010), which do not include M67 and NGC\,6791. On the other hand, the binary fraction in these OCs is on average marginally higher than that in the Solar neighborhood (42\% $\pm$ 9\%; Fischer \& Marcy 1992), as illustrated in Fig.~6. 

GCs may also experience the above dynamical effects as in the OCs, but the relaxation time of GCs is much longer than that of OCs, so that those effects are weaker. 
In fact, three of the four GCs studied here show similar X-ray emissivities with the dEs; only $\omega$ Cen exhibits a significantly lower emissivity, but this cluster is often considered peculiar (Haggard, Cool \& Davies 2009). 
On the other hand, a large number of Galactic GCs are found to have a relatively low binary fraction (Sollima et al.~2007; Milone et al.~2012; Fig.~6), consistent with the general idea that 
dynamical processes in GCs gradually lead to a net destruction of binaries (e.g., Fregeau et al. 2003; van den Berg et al. 2013). 
Of course, the GC X-ray emissivity may also depend on the types of binaries that give rise to the X-ray emission. The differences in the X-ray spectra of the GCs (Table 3), for instance, may point at a different composition of the binary population compared to the field.
A definite answer to whether and how the dynamical effects in GCs would lead to an enhanced or suppressed X-ray emissivity, compared to that in the field, 
awaits study of a sizable sample of the Galactic GCs with sensitive X-ray observations.

The stellar X-ray emissivity may also depend on other properties, such as stellar age and metallicity (Revnivtsev et al. 2008). For instance, the orbital periods and accretion rates of the binary stars evolve with time, which may depend on the stellar age. The metal lines provide a considerable contribution in the $0.5-2$ keV band, which should be scaled with the stellar metallicity. Unfortunately, these relations cannot be well constrained with the currently limited sample.

\begin{figure}
\centering
\includegraphics[width=8.cm]{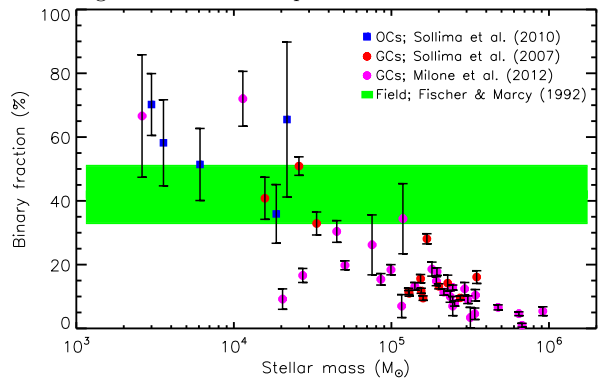}
\caption{
Core binary fraction in globular clusters (Sollima et al. 2007; Milone et al. 2012), and the binary fraction in open clusters (Sollima et al. 2010) and the Solar neighborhood (42\% $\pm$ 9\%; Fischer \& Marcy 1992). 
The stellar mass is estimated from V-band luminosity with $M/L_V$ = 3. None of the OCs and GCs in Fig.~\ref{f:emissivity}, except NGC\,6397, are included in this Figure.
}
\label{f:BF}
\end{figure}

\section{Summary}

We have investigated the unresolved X-ray emission in three Local Group dEs (NGC\,147, NGC\,185 and NGC\,205) using {\sl XMM-Newton} observations.
We have implemented in XSPEC a log-normal plasma temperature distribution model
to fit the spectrum of the unresolved emission, and found similar cumulative X-ray emissivities among the dEs. The averaged X-ray emissivity, presumably originating from old stellar populations such as CVs and ABs, is measured to be $L_{0.5-2\ \rm {keV}}/M_{\ast} = (6.0 \pm 0.5 \pm 1.8) \times 10^{27} \ \rm{erg \ s^{-1} \ M_\odot^{-1}}$. This value is further compared to the cumulative X-ray emissivities of four GCs, of the Solar vicintiy, old OCs and other galaxies from the literature.

We verify previous suggestion that in normal galactic environments with stellar density typical of the field, there exists a quasi-universal X-ray emissivity of old stellar populations. It is important to take into proper account of this ubiquitous stellar component when one attempts to derive the density and temperature distributions of the tenuous, diffuse hot gas in nearby galaxies.

\acknowledgments
We thank M. Revnivtsev for valuable comments and the anonymous referee for suggestions that significantly improve our work.
This work was partially supported by the National Natural Science Foundation of China under grants 11273015 and 11133001, and the National Basic Research Program No.2013CB834905.
C.G. acknowledges support from the program of China Scholarships Council No.201206190041 during his visit to UMass. Z.L. acknowledges support from the Recruitment Program of Global Youth Experts.

\appendix
\twocolumngrid
\section{The log-normal temperature distribution model}
\setcounter{figure}{0}
\renewcommand\thefigure{A\arabic{figure}}
In previous studies (e.g., Revnivtsev et al.2007, 2008), the spectrum of unresolved stellar X-ray emission was typically fitted by a single-temperature model, or a single-temperature plus power-law model. Here we implement in XSPEC a log-normal temperature distribution model. In this model, the distribution of temperature (T) follows a log-normal distribution, i.e., log(T) follows a normal/Gaussian distribution. In astronomy, there are several applications of the log-normal distribution, e.g., the log-normal period distribution of binary system with late-type stars (Kroupa 1995), the log-normal Initial Mass Function (IMF) distribution (Zinnecker 1984), the log-normal star formation history distribution of galaxies (Gladders et al. 2013) and the log-normal distribution of the quiet-Sun FUV continuum (Fontenla et al. 2007).
The log-normal distribution is defined as,
\begin{equation}
f(T)=A \times exp(- \frac{(logT-logT_{peak})^2}{2 \sigma_T ^2}),
\end{equation} 
where A is the normalization, $T_{peak}$ the peak temperature and $\sigma_T$ the dispersion.

The effective temperature of hot plasma in accreting CVs is related with the white dwarf mass (Aizu 1973).
Fig.~A1 shows the mass distribution of CVs from Ritter \& Kolb (2003), which can be fit with a log-normal distribution.
At a first approximation, the temperature distribution is also a log-normal distribution.
Meanwhile, the single-temperature model cannot fit the cumulative stellar emission well (Revnivtsev et al. 2007), due to its multi-temperature nature. Moreover, compared to a two-temperature model, in which at least four parameters (two temperatures and two normalizations) are needed, the log-normal temperature model has only three parameters.
Therefore, the log-normal temperature model is more physical yet simpler.
Fig.~A2 shows the log-normal temperature model with different $\sigma_T$ in comparison with a single-temperature model.

\begin{figure}
\centering
\includegraphics[width=8.5cm,angle=0]{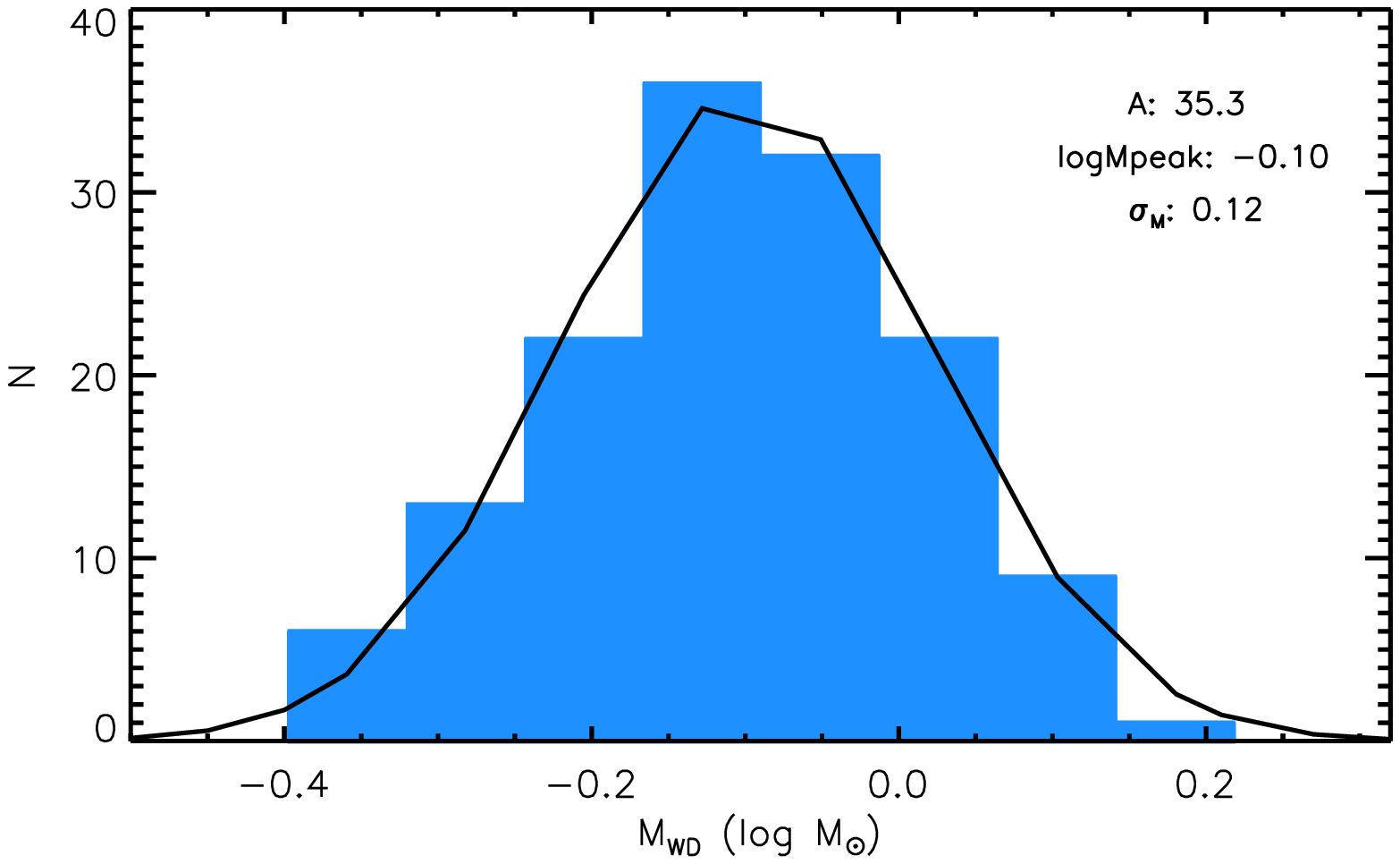}
\caption{
The white dwarf masses distribution of CVs from Ritter \& Kolb (2003), fitted with a log-normal distribution:
$f(M)=A \times exp(- \frac{(logM-logM_{peak})^2}{2 \sigma_M ^2})$.
}
\label{fig:cvmass}
\end{figure}

\begin{figure}
\centering
\includegraphics[width=6.0cm,angle=-90]{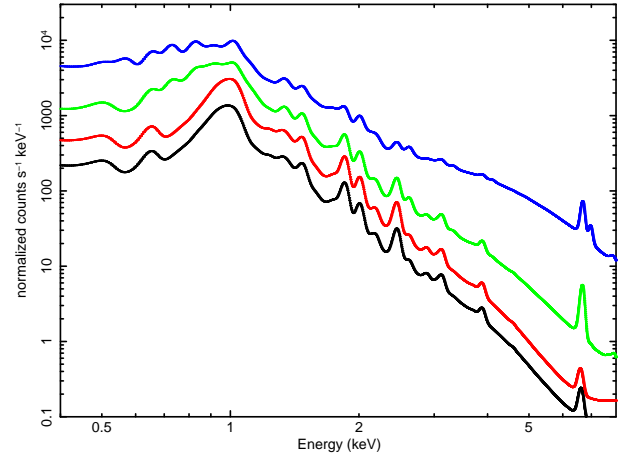}
\caption{A comparison of the single-temperature model (black curve) and the log-normal temperature distribution model with different dispersion $\sigma_T$ (red for $\sigma_T$ = 0.01, green for $\sigma_T$ = 0.5 and blue for $\sigma_T$ = 2), and the same peak temperature $kT_{peak} = 1$ keV. The normalizations are different in order to separate different models.}
\label{fig:multi}
\end{figure}

\end{document}